\documentclass[prl,aps,showpacs,twocolumn,superscriptaddress,floatfix]{revtex4}

\usepackage{graphicx}
\usepackage{epsfig}
\usepackage{amsfonts}
\usepackage{amsmath,amssymb}
\usepackage{bm}

\usepackage[usenames]{color}
\usepackage{ulem}

\def\alphad{$\alpha_{\raisebox{-1pt}{\tiny  D}}$}
\def\rskin{$r_{\rm skin}$}
\def\rn{$r_n$}
\def\rp{$r_p$}
\begin{document}

\title{Electric dipole polarizability and the neutron skin}

\author{J. Piekarewicz}
\affiliation{Department of Physics, Florida State University,
                  Tallahassee, FL 32306, USA}
\author{B. K. Agrawal}
\affiliation{Saha Institute of Nuclear Physics, Kolkata 700064, India}
\author{G. Col\`o}
\affiliation{Dipartimento di Fisica, Universit\`a degli Studi di
  Milano, via Celoria 16, I-20133, Milano, Italy}
\affiliation{INFN,  Sezione di Milano, via Celoria 16, I-20133,
  Milano, Italy}
\author{W. Nazarewicz}
\affiliation{
Department of Physics and Astronomy, University of Tennessee, Knoxville, Tennessee 37996, USA
}%
\affiliation{
Physics Division, Oak Ridge National Laboratory, Oak Ridge, Tennessee 37831, USA
}%
\affiliation{
Institute of Theoretical Physics, University of Warsaw, ul. Ho\.za 69,
PL-00-681 Warsaw, Poland }%
\author{N. Paar}
\affiliation{Physics Department, Faculty of Science, University of Zagreb, Zagreb, Croatia}
\author{P.-G. Reinhard}
\affiliation{Institut f\"ur Theoretische Physik II, Universit\"at Erlangen-N\"urnberg,  Staudtstrasse 7, D-91058 Erlangen, Germany}
\author{X. Roca-Maza}
\affiliation{INFN,  Sezione di Milano, via Celoria 16, I-20133,
  Milano, Italy}
\author{D. Vretenar}
\affiliation{Physics Department, Faculty of Science, University of Zagreb, Zagreb, Croatia}

\begin{abstract}
The recent high-resolution measurement of the electric dipole ($E1$)
polarizability {\alphad} in $^{208}$Pb [Phys. Rev. Lett. {\bf 107},
062502 (2011)] provides a unique constraint on the neutron-skin
thickness of this nucleus. The neutron-skin thickness {\rskin} of
$^{208}$Pb is a quantity of critical importance for our understanding
of a variety of nuclear and astrophysical phenomena. To assess the
model dependence of the correlation between {\alphad} and {\rskin}, we
carry out systematic calculations for $^{208}$Pb, $^{132}$Sn, and
$^{48}$Ca based on the nuclear density functional theory (DFT) using
both non-relativistic and relativistic energy density functionals
(EDFs). Our analysis indicates that whereas individual models exhibit
a linear dependence between {\alphad} and {\rskin}, this correlation
is not universal when one combines predictions from a host of
different models. By averaging over these model predictions, we
provide estimates with associated systematic errors for {\rskin} and
{\alphad} for the nuclei under consideration. We conclude that precise
measurements of {\rskin} in both $^{48}$Ca and ${}^{208}$Pb---combined
with the recent measurement of {\alphad}---should significantly
constrain the isovector sector of the nuclear energy density functional.
\end{abstract}
 

\maketitle

\vspace{1cm}

The Lead Radius Experiment (PREX) \cite{[Hor01a],[PREX05]} at the
Jefferson Laboratory has recently determined the neutron
root-mean-square (rms) radius {\rn} of ${}^{208}$Pb
\cite{[PREX11]}. Parity-violating electron scattering, a powerful
technique used by the PREX collaboration, is particularly sensitive to
the neutron distribution because the neutral weak-vector boson couples
preferentially to the neutrons in the target~\cite{[Don89a]}; the
coupling to the proton is suppressed by the weak mixing angle.  In
spite of the many challenges that it faced, this purely electroweak
measurement may be interpreted with as much confidence as conventional
electromagnetic scattering experiments that have been used for decades
to accurately map the electric charge distribution of the nucleus.

A quantity that is related to the neutron radius is the neutron-skin
thickness {\rskin}={\rn}$-${\rp}, namely, the difference between the
rms neutron and proton radii. The importance of the neutron skin lies
in its strong sensitivity to the poorly known isovector density
$\rho_1\!=\!\rho_n\!-\!\rho_p$. Given that {\rskin} is a strong
indicator of isovector properties, the determination of {\rn} of a
heavy nucleus is a problem of fundamental importance with far-reaching
implications in areas as diverse as nuclear structure
\cite{[Bro00a],[Fur02],[Cen10],[Rei10]}, atomic parity
violation~\cite{[Pol92]}, and neutron-star
structure~\cite{[Hor01],[Ste05]}.  By measuring the neutron form
factor of $^{208}$Pb at a moderate momentum transfer of
$q\!\approx\!0.475\,{\rm fm}^{-1}$, and through an extrapolation to
low-momentum transfers~\cite{[Fur02],[Roc11]}, PREX was able to
determine the following values for the neutron radius and neutron-skin
thickness: {\rn}=$5.78^{+0.16}_{-0.18}$\,fm and
{\rskin}=$0.33^{+0.16}_{-0.18}$\,fm \cite{[PREX11]}.

Another observable that is a strong indicator of isovector properties
is the electric dipole polarizability {\alphad} related to the
response of the nucleus to an externally applied electric field.  For
stable medium-to-heavy nuclei with a moderate neutron excess, the
dipole response is largely concentrated in the giant dipole resonance
(GDR) of width 2-4 MeV that exhausts almost 100\% of the
energy-weighted sum rule~\cite{[Har01]}.  For this isovector mode of
excitation -- perceived as an oscillation of neutrons against
protons -- the symmetry energy $a_\mathrm{sym}$ acts as the restoring
force. Models with a soft symmetry energy, namely, those that change
slowly with density, predict larger values for $a_\mathrm{sym}$ at the
lower densities of relevance to the excitation of this mode
\cite{[Rei99d],[Tri08]}.  In this context, the inverse energy-weighted
$E1$ sum rule $m_{-1}$ -- a quantity directly proportional to
{\alphad} -- is of particular interest as it is highly sensitive to the
density dependence of the symmetry energy.  This sensitivity suggests
the existence of a correlation: the larger {\rskin}, the larger
{\alphad}. Indeed, the approximate proportionality of these two
quantities is expected based on both macroscopic
arguments \cite{[Lip82],[Sat06]} and microscopic calculations
\cite{[Rei10],[Pie11]}. The recently completed
high-resolution $(\vec{p},\vec{p}^{\,\prime})$ measurement at RCNP of the
distribution of $E1$ strength in ${}^{208}{\rm Pb}$ over a wide range
of excitation energy~\cite{[Tam11]} has, therefore,  created considerable
excitement. Of particular relevance to our work is the precise value
of the measured electric dipole polarizability of ${}^{208}{\rm Pb}$:
{\alphad}=(20.1$\pm$0.6)\,fm$^3$.

It is the purpose of this work to examine possible correlations
between the dipole polarizability and the neutron-skin thickness of
${}^{208}{\rm Pb}$. Generally, to assess a linear correlation between two
observables $A$ and $B$ within {\it one given model}, one resorts to a
least-squares covariance analysis, with the correlation coefficient
\begin{equation}
{C}_{AB} =
\frac{|\overline{\Delta A\,\Delta B}|}
    {\sqrt{\overline{\Delta A^2}\;\overline{\Delta B^2}}},
\label{correlator}
\end{equation}
 providing the proper statistical measure~\cite{[Bra97a]}. 
In Eq.~(\ref{correlator}) the overline means an average over the
statistical sample. A
value of $|{C}_{AB}|=1$ means that the two observables are fully
correlated whereas ${C}_{AB}=0$ implies that they are totally
uncorrelated. Recently, the statistical measure ${C}_{AB}$ was used to
study correlations between various nuclear observables \cite{[Rei10]}
in the context of the Skyrme SV-min model \cite{[Klu09]}. In
particular, it was concluded that good isovector indicators that
strongly correlate with the neutron radius of $^{208}$Pb are its
electric dipole polarizability as well as neutron skins and radii of
neutron-rich nuclei~\cite{[Rei10]}.  Indeed, by relying on the strong
correlation between {\alphad} and {\rskin} (${C}_{AB}$=0.98) predicted
by such DFT calculations, Tamii {\it et al.}  deduced a value of
0.156$^{+0.025}_{-0.021}$~fm for the neutron-skin thickness of
$^{208}$Pb. 

However, the correlation coefficient ${C}_{AB}$ cannot 
assess systematic errors that reflect constraints and limitations 
of a given model~\cite{[Rei10]}. Such systematic uncertainties can 
only emerge by comparing different models (or sufficiently flexible 
variants of a model) and this is precisely what has been done in this 
Letter. To assess the linear dependence between two observables 
$A$ and $B$ for a sample of {\it several models}, the correlation 
coefficient ${C}_{AB}^{\rm models}$ is now obtained by averaging over
the predictions of those models. Although the correlation coefficient
${C}_{AB}^{\rm models}$ determined in such a way may not have a clear
statistical interpretation, it is nevertheless an excellent indicator of 
linear dependence.

\begin{figure}[htb]
\includegraphics[width=0.5\textwidth]{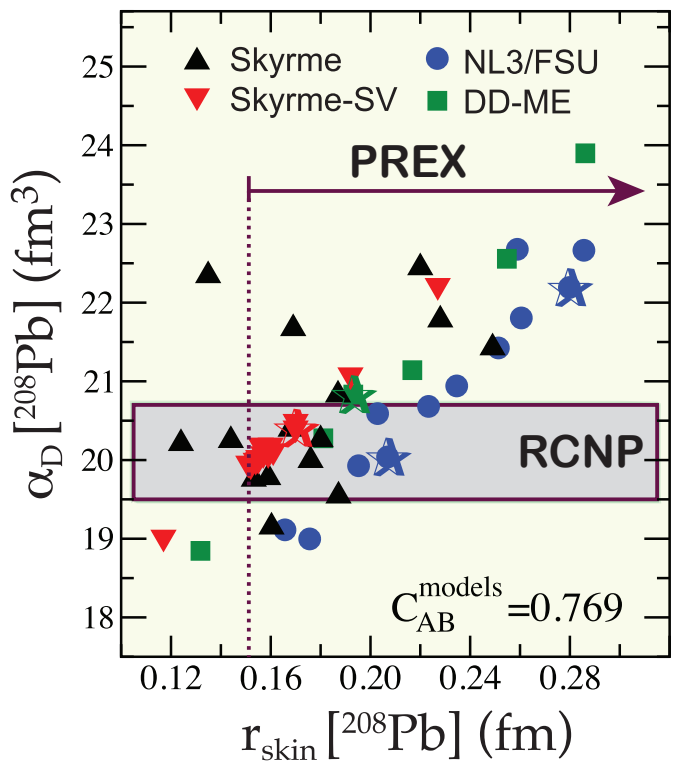}
\caption{(Color online) Predictions from 48 nuclear EDFs
  discussed in the text for the electric dipole polarizability 
 and neutron-skin thickness of ${}^{208}$Pb. Constrains on 
 the neutron-skin thickness from PREX~\cite{[PREX11]} and 
 on the dipole polarizability from RCNP~\cite{[Tam11]} have 
 been incorporated into the plot.}
\label{Fig1}
\end{figure}
\vspace{-0.05cm}
To this end, we have computed the distribution of $E1$ strength using
both relativistic and non-relativistic DFT approaches with different
EDFs.  In all cases, these self-consistent models have been calibrated
to selected global properties of finite nuclei and some parameters of
nuclear matter.  Once calibrated, these models are used without any
further adjustment to compute the $E1$ strength
$R_{\raisebox{-1pt}{\tiny E1}}$ using a consistent random-phase
approximation. The electric dipole polarizability is then obtained 
from the inverse energy-weighted sum \cite{[Lip89],[Rei10],[Pie11]}:
\begin{equation}
 \alpha_{\raisebox{-1pt}{\tiny D}} = \frac{8\pi}{9}e^{2}
  \int_{0}^{\infty}\!\omega^{-1} 
   R_{\raisebox{-1pt}{\tiny E1}} (\omega)\,d\omega \;.
\label{AlphaD}
\end{equation}

The relation between {\alphad} and {\rskin} for $^{208}$Pb is
displayed in Fig.~\ref{Fig1} using the predictions from the 48 EDFs
chosen in this work. In particular, the up-triangles mark predictions
from a broad choice of Skyrme EDFs that have been widely used in the
literature: SGII, SIII, SkI3, SkI4, SkM$^*$, SkO, SkP, SkX, SLy4,
SLy6, (see Refs.~\cite{[Ben03],[Rei06]} for the original references),
Sk255 \cite{[Agr03]}, BSk17 \cite{[Gor09]}, LNS \cite{[Cao06a]}, and
UNEDF0 and UNEDF1 \cite{[Kor10b]}. In addition, we consider a
collection of relativistic and Skyrme EDFs that have been
systematically varied around an optimal model without a significant
deterioration in the quality of the fit. (This is particularly true for
the case of the isovector interaction which at present remains poorly
constrained.)  Those results are marked in Fig.~\ref{Fig1} as
NL3/FSU~\cite{[Agr10],[Pie11]} (circles), DD-ME~\cite{[Vre03]}
(squares), and Skyrme-SV~\cite{[Klu09]} (down-triangles).  Note that
the ``stars'' in the figure are meant to represent the predictions
from the optimal models within the chain of systematic variations of
the symmetry energy. At first glance a clear (positive) correlation
between the dipole polarizability and the neutron skin is discerned.

Yet on closer examination, one observes a significant scatter in
the results, especially for the standard Skyrme models. In particular,
by including the predictions from all the 48 EDFs considered here, the
correlation ${C}_{AB}^{\rm models}$=0.77 is obtained.  However, as
seen in Table~\ref{Table1}, within each set of the systematically
varied models an almost perfect correlation is found.  Note that
by imposing the recent experimental constraints on {\rskin} and
{\alphad}, several of the models -- especially those with either a
very soft or very stiff symmetry energy -- may already be ruled out.
Thus, if we average our theoretical results over the set of 25 EDFs
(``Set-25'') whose predictions fall within the RCNP value of
{\alphad}, we obtain {\rskin}=(0.168$\pm$0.022)\,fm, a value that is
fairly close to the one obtained in Ref.~\cite{[Tam11]}.  It is to be noted that
23 of those 25 EDFs are consistent with the PREX
constraint of {\rskin} greater than 0.15\,fm. However, the average
theoretical value is significantly below the current PREX mean of
0.33\,fm \cite{[PREX11]}. If confirmed by the anticipated
higher-precision (1\%) PREX run, this large difference could either
indicate the need for significant revisions of current nuclear
structure models or of the models employed by PREX to deduce {\rskin}
from the neutron form factor, or both. Provided that the new PREX and
theoretical average values of {\rskin} are closer, in order to
discriminate between theoretical models of Fig.~\ref{Fig1} and further
constraint theory, an accuracy of at least 0.03\,fm on the
experimental value of the neutron radius will be required. Based on
the central PREX value of {\rn}=5.78\,fm \cite{[PREX11]}, this
translates to a 0.5\% measurement.

\begin{table*}[!]
\begin{tabular}{|l|ccc|ccc|ccc|}
 \hline
 &
 \multicolumn{3}{|c|}{\rule[-2mm]{0mm}{6mm}
  $\alpha_{\raisebox{-1pt}{\tiny D}} [{}^{208}{\rm Pb}]$}&
 \multicolumn{3}{|c|}{\rule[-2mm]{0mm}{6mm}
  {\rskin}[$^{132}$Sn]} &
 \multicolumn{3}{|c|}{\rule[-2mm]{0mm}{6mm}
 {\rskin}[$^{48}$Ca]}\\
\hline
  Model 
  & $C_{AB}^{\rm model}$ & Slope(fm$^2$) & Intercept(fm$^3$)
  & $C_{AB}^{\rm model}$ & Slope & Intercept(fm) 
  & $C_{AB}^{\rm model}$ & Slope & Intercept(fm)  \\
\hline
 Skyrme   & 0.996 & 29.08 & 15.53&
                    0.999 &   1.06 &   0.06 &
                    0.977 &   0.60 &   0.08\\
 DD-ME    & 0.994 & 31.99 &  14.52&
                    1.000 &   1.06 &    0.05&
                    1.000 &   0.53 &    0.08\\
 NL3/FSU  & 0.994 & 29.89 &  13.97 &
                     1.000 &   1.04 &    0.05&
                     0.987 &   0.50 &    0.09\\
\hline
\end{tabular}
\caption{Least-square correlation coefficient, slope, and
  intercept between various observables and the 
 neutron-skin thickness of ${}^{208}$Pb for the  systematically varied
 models: NL3/FSU, DD-ME, and Skyrme-SV. Slope and intercept are obtained by fitting a straight
 line through the data.}
\label{Table1}
\end{table*}

Using either lighter nuclei measured at larger momentum transfers or
nuclei with a larger neutron excess will increase the parity-violating
asymmetry. Therefore, it is pertinent to ask whether parity-violating
experiments in other nuclei may be warranted~\cite{[Ban11]}.  To this
end, we have computed data-to-data relations between the neutron-skin
thickness of ${}^{208}$Pb and the neutron-skin thickness of two doubly
magic neutron-rich nuclei: stable ${}^{48}$Ca and unstable
${}^{132}$Sn. While parity-violating experiments on radioactive nuclei
are unlikely to happen in the foreseeable future, such experiments on
stable targets may serve to calibrate experiments with hadronic probes
that could eventually be used to extract neutron radii of short-lived
systems such as ${}^{132}$Sn.
\begin{figure}[ht]
\includegraphics[width=0.45\textwidth]{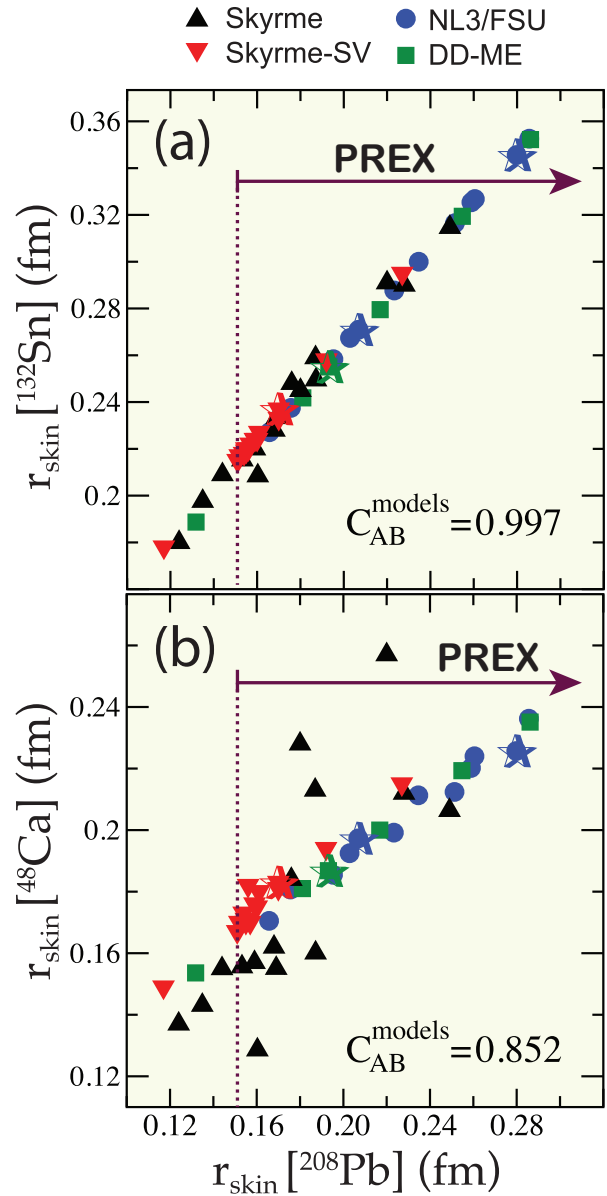}
\caption{(Color online) Predictions from the 48 nuclear EDFs used
 in the text for the neutron-skin thickness of ${}^{208}$Pb and 
 ${}^{132}$Sn (a) and ${}^{48}$Ca (b). Constrains on the neutron-skin 
 thickness from PREX~\cite{[PREX11]} have been incorporated into 
 the plot.}
\label{Fig2}
\end{figure}

Figure~\ref{Fig2}a displays model predictions for the neutron-skin
thickness of ${}^{132}$Sn as a function of the corresponding {\rskin}
in ${}^{208}$Pb. The displayed correlation is both strong and fairly
model independent. Indeed, ${C}_{AB}^{\rm models}$=0.997 for the set
of 48 EDFs used in this work, and it is even closer to unity for the
systematically varied forces listed in Table~\ref{Table1}.  This
suggests that new experimental information on {\rskin} in ${}^{132}$Sn
is not likely to provide additional constraints on the theoretical models
used here, provided that an accurate measurement of the neutron-skin
thickness of ${}^{208}$Pb is available. Averaging our results, a
theoretical estimate for {\rskin} in ${}^{132}$Sn of
(0.232$\pm$0.022)\,fm is obtained with Set-25. In addition, we predict a value of
(10.081$\pm$0.150)\,fm$^3$ for {\alphad}.

The situation for the case of the neutron-skin thickness in
${}^{48}$Ca shown in Fig.~\ref{Fig2}b is different. Whereas the
correlation coefficient among the three systematically varied models
 remains close to unity (see
Table~\ref{Table1}) there is a significant spread in the predictions
of all 48 models that is driven primarily by the traditional Skyrme
forces. This suggests that an accurate measurement of {\rskin} in
${}^{208}$Pb is not sufficient to significantly constrain {\rskin} in
${}^{48}$Ca.  Conversely, by measuring the neutron-skin thickness of
both ${}^{48}$Ca and ${}^{208}$Pb, and incorporating the recent
measurement of {\alphad} in ${}^{208}$Pb, one should be able to
significantly constrain the isovector sector of the nuclear EDF.  The
theoretical model-averaged estimate for {\rskin} in ${}^{48}$Ca is
(0.176$\pm$0.018)\,fm for Set-25. Moreover, a prediction of 
(2.306$\pm$0.089)\,fm$^3$ for 
{\alphad} in $^{48}$Ca is obtained.

In summary, we have examined the correlation between the electric
dipole polarizability and neutron-skin thickness of ${}^{208}$Pb using
a large ensemble of 48 reasonable nuclear energy density functionals.
Physical arguments based on a macroscopic analysis suggest that these
two isovector observables should be correlated, although this
correlation may display some systematic model dependence. In fact, we
have found that as accurately calibrated models are systematically
varied around their optimal value, strong correlations between
{\rskin} and {\alphad} in $^{208}$Pb do emerge.  As these models are
combined, however, the correlation weakens.  To study the associated
systematic errors, we have performed calculations of {\alphad} and
{\rskin} using the subset of models that are consistent with the
experimental value of {\alphad} in $^{208}$Pb~\cite{[Tam11]}. Using
this subset we predict the following ``model-averaged" values of
{\rskin}: (0.168$\pm$0.022)\,fm in $^{208}$Pb, (0.232$\pm$0.022)\,fm
in $^{132}$Sn, and (0.176$\pm$0.018)\,fm in $^{48}$Ca---as well as an
electric dipole polarizability of: (10.081$\pm$0.150)\,fm$^3$ in
$^{132}$Sn and (2.306$\pm$0.089)\,fm$^3$ in $^{48}$Ca.  Given these
results, we conclude that the follow-up PREX measurements of {\rskin}
in $^{208}$Pb will be of great value in further constraining the
poorly known isovector sector of the nuclear EDF.  Moreover, the
analysis carried out in this work has enabled us to identify
additional critical observables that could help discriminate among
theoretical models.  Specifically, we endorse a measurement of the
neutron radius in $^{48}$Ca, as it provides information that is
complimentary to the $^{208}$Pb measurement.  Finally, in the near
future we aim to present a complementary study of {\rskin}, {\alphad},
and the low-energy $E1$ strength by means of a detailed statistical
covariance analysis within the realm of accurately calibrated
models~\cite{[Rei10]}.

Useful discussions with Chuck Horowitz are gratefully
acknowledged. This work was supported in part by the Office of Nuclear
Physics, U.S. Department of Energy under Contract
Nos. DE-FG05-92ER40750 (FSU), DE-FG02-96ER40963 (UTK); and by the BMBF
under Contract 06ER9063.

\bibliographystyle{unsrt}
\bibliography{alphad} 

\end{document}